\documentclass[final,1p,times]{elsarticle}

\usepackage{amssymb}
\usepackage{amsmath}
\usepackage{textcomp}
\usepackage{multirow}
\usepackage{makecell}
\usepackage{bm}
\usepackage{siunitx}
\usepackage{threeparttable}
\usepackage{placeins}
\usepackage[colorlinks=true,citecolor=blue]{hyperref}
\usepackage{lineno}
\newcommand{\matr}[1]{\mathbf{#1}}
\def\nuc#1#2{\relax\ifmmode{}^{#1}{\protect\text{#2}}\else${}^{#1}$#2\fi}

\journal{Nuclear Instruments and Methods A}

\usepackage{array}
\newcolumntype{C}[1]{>{\centering\arraybackslash}m{#1}}

\begin{document}
\begin{frontmatter}

\title{Ultra-sensitive radon assay using an electrostatic chamber in a recirculating system}

\author[36]{A.~Anker \fnref{fn8}}
\author[36]{P.~A.~Breur}
\author[36]{B.~Mong \fnref{fn9}}
\author[5]{P.~Acharya}
\author[1,38]{A.~Amy}
\author[32]{E.~Angelico}
\author[29]{I.~J.~Arnquist}
\author[14]{A.~Atencio}
\author[1]{J.~Bane}
\author[27]{V.~Belov}
% ...

\address[36]{SLAC National Accelerator Laboratory, Menlo Park, CA 94025, USA}
\address[5]{Department of Physics and Astronomy, University of Alabama, Tuscaloosa, AL 35405, USA}
\address[1]{Amherst Center for Fundamental Interactions and Physics Department, University of Massachusetts, Amherst, MA 01003, USA}
\address[38]{SUBATECH, Nantes Universit\'e, IMT Atlantique, CNRS/IN2P3, Nantes 44307, France}
\address[32]{Physics Department, Stanford University, Stanford, CA 94305, USA}
\address[29]{Pacific Northwest National Laboratory, Richland, WA 99352, USA}
\address[14]{Department of Physics, Drexel University, Philadelphia, PA 19104, USA}
\address[27]{National Research Center ``Kurchatov Institute'', Moscow, 123182, Russia}
% ...

\fntext[fn8]{aanker@slac.stanford.edu}
\fntext[fn9]{bung@slac.stanford.edu}
\fntext[fn1]{Also at: University of Chinese Academy of Sciences, Beijing, China}
\fntext[fn2]{Also at: Facility for Rare Isotope Beams, Michigan State University, East Lansing, MI 48824, USA}
\fntext[fn3]{Also at: Center for Energy Research and Development, Obafemi Awolowo University, Ile-Ife, 220005 Nigeria}
\fntext[fn4]{Now at: Canon Medical Research USA, Inc.}
\fntext[fn5]{Now at: Bates College, Lewiston, ME 04240, USA}

\begin{abstract} 
Rare event searches such as neutrinoless double beta decay and Weakly Interacting Massive Particle detection require ultra-low background detectors.
Radon contamination is a significant challenge for these experiments, which employ highly sensitive radon assay techniques to identify and select low-emission materials. 
This work presents the development of ultra-sensitive electrostatic chamber (ESC) instruments designed to measure radon emanation in a recirculating gas loop, for future lower background experiments. 
Unlike traditional methods that separate emanation and detection steps, this system allows continuous radon transport and detection. 
This is made possible with a custom-built recirculation pump. A Python-based analysis framework, PyDAn, was developed to process and fit time-dependent radon decay data. 
Radon emanation rates are given for various materials measured with this instrument. A radon source of known activity provides an absolute calibration, enabling statistically-limited minimal detectable activities of 20~$\mu$Bq. 
These devices are powerful tools for screening materials in the development of low-background particle physics experiments. 
\end{abstract}

\end{frontmatter}

\section{Introduction}
Rare event searches like those for neutrinoless double beta decay ($0\nu\beta\beta$) and Weakly Interacting Massive Particles (WIMPs) require large exposure and extremely low backgrounds to be sensitive to a few potential events per year. 
To detect these rare events, many of the most sensitive experiments utilize liquid xenon (LXe) time projection chambers (TPCs), including experiments such as EXO-200 \cite{Auger2012}, LZ~\cite{LZ_general} and XENONnT \cite{XENONnT_general}, as well as proposed experiments like nEXO~\cite{nEXO_PCDR}, XLZD~\cite{XLZD_0bb}, and PandaX-xT \cite{Abdukerim2024}.
One of the major background sources common to all of these experiments is from radon dissolved in the LXe. 
Existing experiments have achieved radon concentrations on the order of $1~\mu$Bq/kg~\cite{xenonnt_rn,LZ_Rn,Albert2015}, while future experiments will require closer to $0.1~\mu$Bq/kg~\cite{XLZD_Rn,nEXO_Rn,PandaX-xT_Rn}. 
Meeting these stringent background goals requires careful screening of all xenon-wetted components to select materials that produce as little radon as possible.
\subsection{Radon as a Background Source}
\begin{figure*}[h]
\centering
  \includegraphics[width=0.8\textwidth]{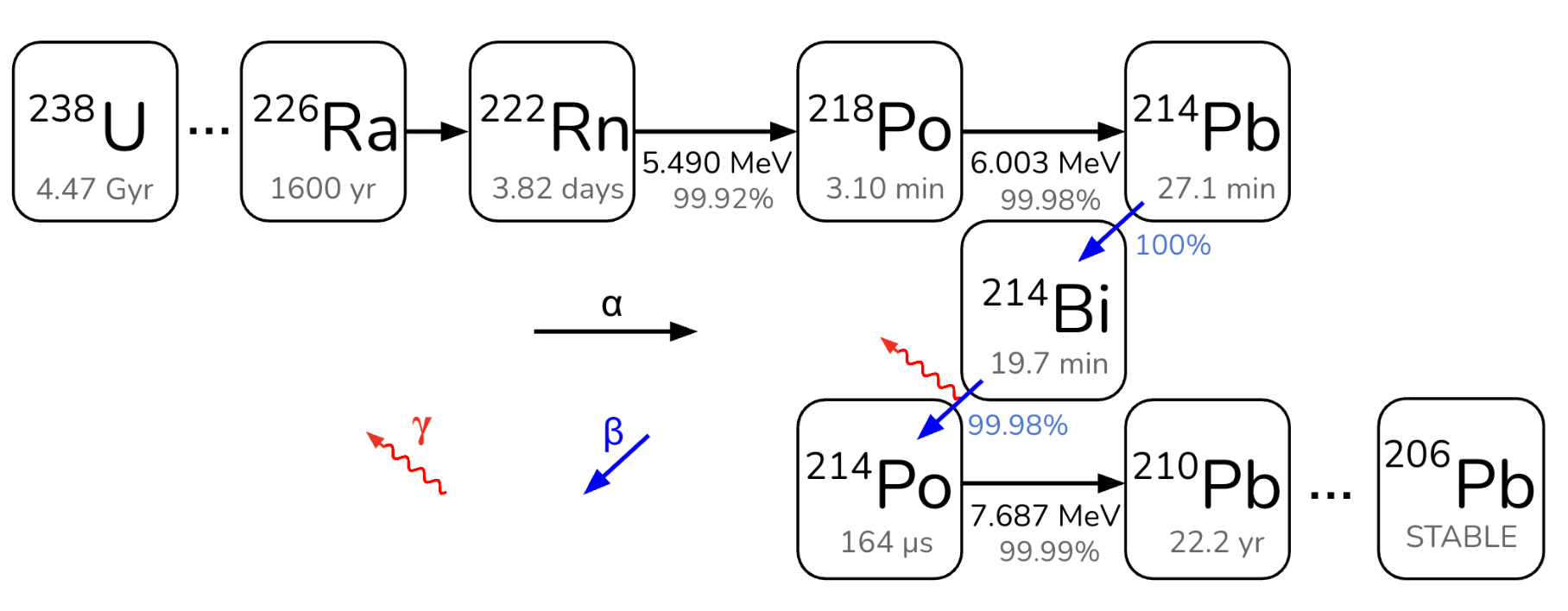}
  \caption{Condensed diagram of the $^{238}\text{U}$ decay series. The half-lives and branching ratios for isotopes of interest are given as well as the alpha energy values \cite{ensdf}.}
  \label{fig:decay_chain}
\end{figure*}

In these low background experiments, radon is produced by the construction materials containing certain primordial radionuclides. 
If a decay producing radon occurs near the surface, it can be emitted from the material through both nuclear recoil and diffusion, the combination of which is called emanation.
This work uses the generally adopted unit of Becquerel (Bq) to express the rate of emanation, i.e. the number of decays that results in an emanated radon atom per second.
There are three natural decay series containing radon: $^{235}\text{U}$, $^{232}\text{Th}$, and $^{238}\text{U}$ containing $^{219}\text{Rn}$ ($T_{1/2}=3.96~\mathrm{s}$), $^{220}\text{Rn}$ ($T_{1/2}=55.6~\mathrm{s}$), and $^{222}\text{Rn}$ ($T_{1/2}=3.82~\mathrm{d}$) respectively. 
The former two are not significant background sources due to their short half-lives, which limit the available diffusion time to emanate.
The radon that does emanate will decay quickly into daughters that tend to plate out before mixing into the xenon~\cite{LZ_RBC_Program}. 
The primary background concern is from $^{222}\text{Rn}$ progeny, whose decay chain is summarized in Fig.~\ref{fig:decay_chain}. 
For $0\nu\beta\beta$ searches, the background comes from the $\beta$-decay daughter of $^{214}\text{Bi}$, which can produce a $\gamma$-ray near the $Q$ value in $^{136}\text{Xe}$ ($Q_{\beta\beta}= 2458.10\pm0.31~\mathrm{keV}$) \cite{q_val1}. 
For dark matter searches, the main background is the $\beta$ decay of $^{214}\text{Pb}$, which can produce single-site events into the WIMP search energy range ($<10$~keV). 

\subsection{Radon Assay Overview}
Different approaches to performing radon assay exist, which fall into two broad categories based on their measurement process.
The first type, which we refer to as injection systems, separate the emanation and measurement process into two steps by first emanating radon in a closed volume, and then transferring that radon to a detector to be counted.
The second type, which we refer to as recirculation systems, perform the emanation and measurement in the same closed volume, typically in separate chambers connected in a recirculation loop with a gas pump. 
Each measurement type has advantages and disadvantages, which are briefly discussed below. 

The injection technique has multiple examples, each utilizing a different detector. 
In all cases the process begins by placing a sample in a closed volume to emanate radon.
One technique involves dissolving the $^{222}\text{Rn}$ in a liquid scintillator and then using photomultiplier tubes (PMTs) to measure the time-correlated decay of $^{214}\text{Bi}$ and $^{214}\text{Po}$ progeny \cite{sazzad24_rn}. 
Other techniques separate the radon from other gases using cryogenic traps, so that the radon can be injected into $\alpha$-counters:
e.g. Lucas cells coated with scintillating material counted with PMTs \cite{Liu1993, LucasCell_Blevis2004}, or high purity quartz proportional counters \cite{Zuzel2009,RnAssays_XENON1T}.
Concentrating the radon has an advantage that the detector can be small with very low background and still be capable of assaying large volumes.
For samples contaminated with xenon where the injection gas volume could not be sufficiently concentrated, ESCs have been used \cite{brunner_thesis,RnAssays_XENON1T}. 
The downside of the single injection technique is that the sensitivity is limited by the transfer and counting of a small number of decaying $^{222}\text{Rn}$ atoms ($10~\text{atoms}$ corresponds to $\sim 20~\mathrm{\mu Bq}$).
These low count rates result in large statistical uncertainties unless repeated measurements are combined, such as in Ref. \cite{Armstrong2023}. 
The sensitivity is further limited by the backgrounds of the emanation chambers~\cite{Zuzel2009}.

\begin{figure*}[h]
    \centering
    \includegraphics[width=0.8\linewidth]{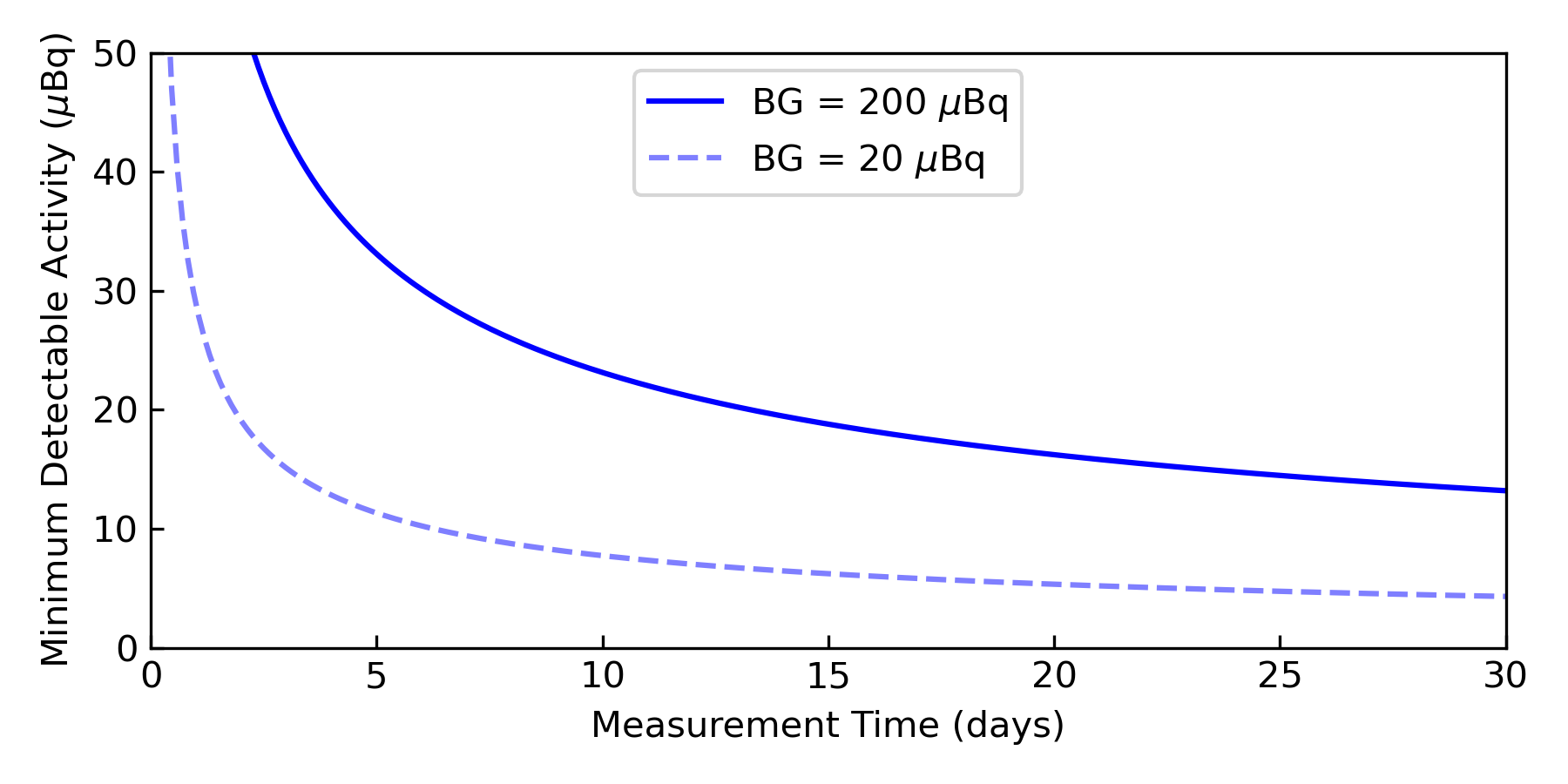}
    \caption{The minimum detectable activity of the recirculation instrument with 68\% confidence level. Assumptions are that 40\% of all radon decays result in a $^{214}\text{Po}$ count, and the recirculation loop volume is much less than the detector volume. The recirculation method, after some stabilization time, measures a rate proportional to the radon emanation of the sample and thus becomes more sensitive over time. This figure is modified from \cite{brunner_thesis,Mamedov2011}}
    \label{fig:recirc_inject}
\end{figure*}
In the recirculation system, a sample emanates radon in an external chamber which is connected to the ESC chamber with a constant flow of carrier gas by a recirculation pump.
These instruments were first developed for the SNO water $^{226}\text{Ra}$-assay systems \cite{SNO_ESC_NIM,Andersen2003,Farine2005}. 
Since the sample is continuously emanating radon, the detector will eventually measure counts at a rate proportional to the emanation rate, after an initial period during which the populations reach steady state.
The time dependence of the observed rate distinguishes the $^{226}\text{Ra}$ supported fractions from the $^{222}\text{Rn}$ that is initially present.
Additionally, the radon can be transported to the ESC by the carrier gas quickly ($< 1~\mathrm{sec}$), so this technique can be sensitive to all three radon isotopes. 
One downside of this technique is that the system has additional radon emitting surfaces in the recirculation loop, and thus larger backgrounds.
Secondly, it cannot detect radon that decays outside of the ESC vessel, so it is less efficient for larger emanation chambers.

\subsection{Sensitivity of the Recirculation ESC}
The nEXO collaboration supported the development of a new generation of sensitive radon assay instruments to screen all of their xenon-wetted construction materials. 
In this work, we describe these instruments, which utilize electrostatic chamber (ESC) detectors that remain connected to the sample via a recirculating gas loop.
The sensitivity target of these measurements is driven by the design goal of having less than $1.26~\mathrm{m Bq}$ of $^{222}\text{Rn}$ in the entire $5000~\mathrm{kg}$ of xenon (less than $0.252~\mu$Bq/kg)~\cite{nEXO_PCDR,nEXO_Rn}. 
Figure~\ref{fig:recirc_inject} shows the 68\% CL (confidence level) minimal detectable activity (MDA) of the recirculating instrument with two background scenarios: $200~\mathrm{\mu Bq}$ which represents the current instruments, and $20~\mathrm{\mu Bq}$ which is equivalent to the best detector-only backgrounds achieved by the injection techniques~\cite{Zuzel2009,RnAssays_XENON1T}. 
While achieving backgrounds in the recirculating system as low as $20~\mathrm{\mu Bq}$ is unrealistic, at long measurement times the current instruments (with a background of $200~\mathrm{\mu Bq}$) can theoretically achieve better than $\sim 20~\mathrm{\mu Bq}$ MDA.
After considering all of the measurement systematics (see Sec.~\ref{sec:measurements}) the instrument has demonstrated near theoretical performance.

In the following section (Sec.~\ref{sec:Overview}) the hardware, electronics, and data acquisition of the recirculation instruments are described. 
In Section~\ref{sec:PyDAn} an analysis process that fits the counts to the Bateman equation solutions is outlined.
Finally, Section~\ref{sec:measurements} discusses the instrument calibration and systematics that relate the fits to an emanation rate, and a table of emanation results are presented.

\section{The Recirculation ESC for Radon Assay}
\label{sec:Overview}
\subsection{Recirculation System Overview}
\begin{figure*}
\centering
  \includegraphics[width=0.8\textwidth]{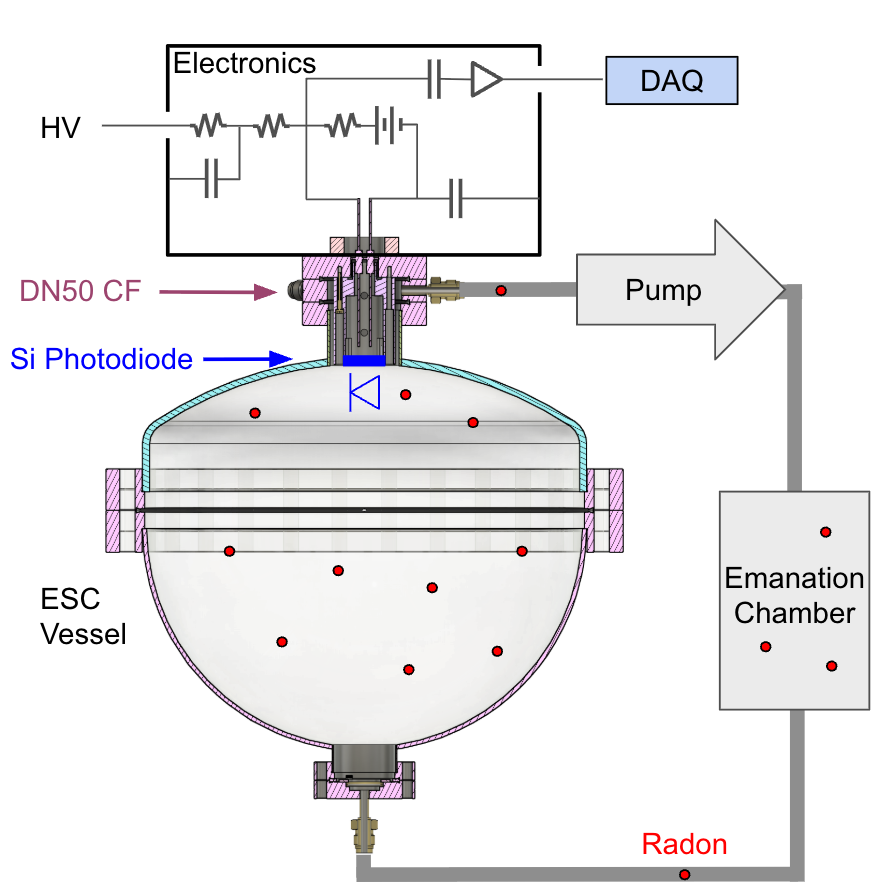}
  \caption{Simplified diagram of the radon assay system. A pump recirculates the carrier gas, transporting radon from the emanation source to the ESC vessel where radon progeny can be collected and detected. The electronics circuit housed above the ESC provides the negative bias voltage to the Si photodiode detector to establish the drift-field inside the grounded chamber. It also provides the necessary circuitry for readout of charge on the blocking capacitor (BC) created by $\alpha$ particles entering the Si photodiode.}
  \label{fig:esc_recir}
\end{figure*}

The recirculation assay system built at the SLAC National Accelerator Laboratory consists of a closed loop in which the emanation sample and ESC are connected as shown in Fig.~\ref{fig:esc_recir}. 
The system comprises an ESC vessel, a sample emanation chamber, and a gas recirculation pump, all connected using ultra-high-vacuum-compatible plumbing and all-metal seals. 
The system is filled with a carrier gas of high purity (99.999\%) argon or nitrogen to a pressure of 1~atm, which is then continuously recirculated to transport radon from the emanation chamber to the ESC during the measurement. 
When radon decays inside the ESC vessel, positive daughter ions are electrostatically drifted to the detector, which is biased at negative high voltage relative to the grounded vessel. 
The detection of the subsequent $\alpha$-decays enables the extraction of the overall radon activity in the gas, and therefore, the emanation rate of the sample.

\subsection{System Hardware}
The ESC vessel is constructed from two stainless steel pressure vessel heads, a hemispherical bottom and a 2:1 flanged and dished top.
The top head contains a DN50 CF vacuum flange for the detector mount, as well as ports for a pressure transducer and the gas outlet.
The bottom head contains the gas inlet and a sintered metal filter to prevent particulate from getting into the ESC.
This filter also plates out radon daughters produced outside of the ESC vessel, which is important for the volume sharing correction discussed in Sec.~\ref{sec:measurements}.
The vessel shape is expected to efficiently drift ions to the detector within the entire volume, while being made from off the shelf parts.
The detector is a Hamamatsu S3204-09 large area (18 x 18 mm) Si PIN photodiode, mounted on a two conductor high-voltage CF feedthrough using a custom PTFE holder.  
Two of these systems are actively taking data: a $25.4~\text{cm}$ ($10~\text{inch}$) diameter ESC with a $7.12~\text{liter}$ volume, and a $30.5~\text{cm}$ ($12~\text{inch}$) diameter ESC with a $11.9~\text{liter}$ volume. 
However, data presented in this work is exclusively from the $7.12~\text{liter}$ ESC.
The vessels' tank heads were mechanically polished and then electropolished before construction.
Electropolishing is expected to reduce the emanation background of the vessel walls by removing intrinsic radium near the surface and by reducing the surface area \cite{electropolishing_reduce_Rn}. 

\begin{figure*}[t]
\centering
  \includegraphics[width=0.8\textwidth]{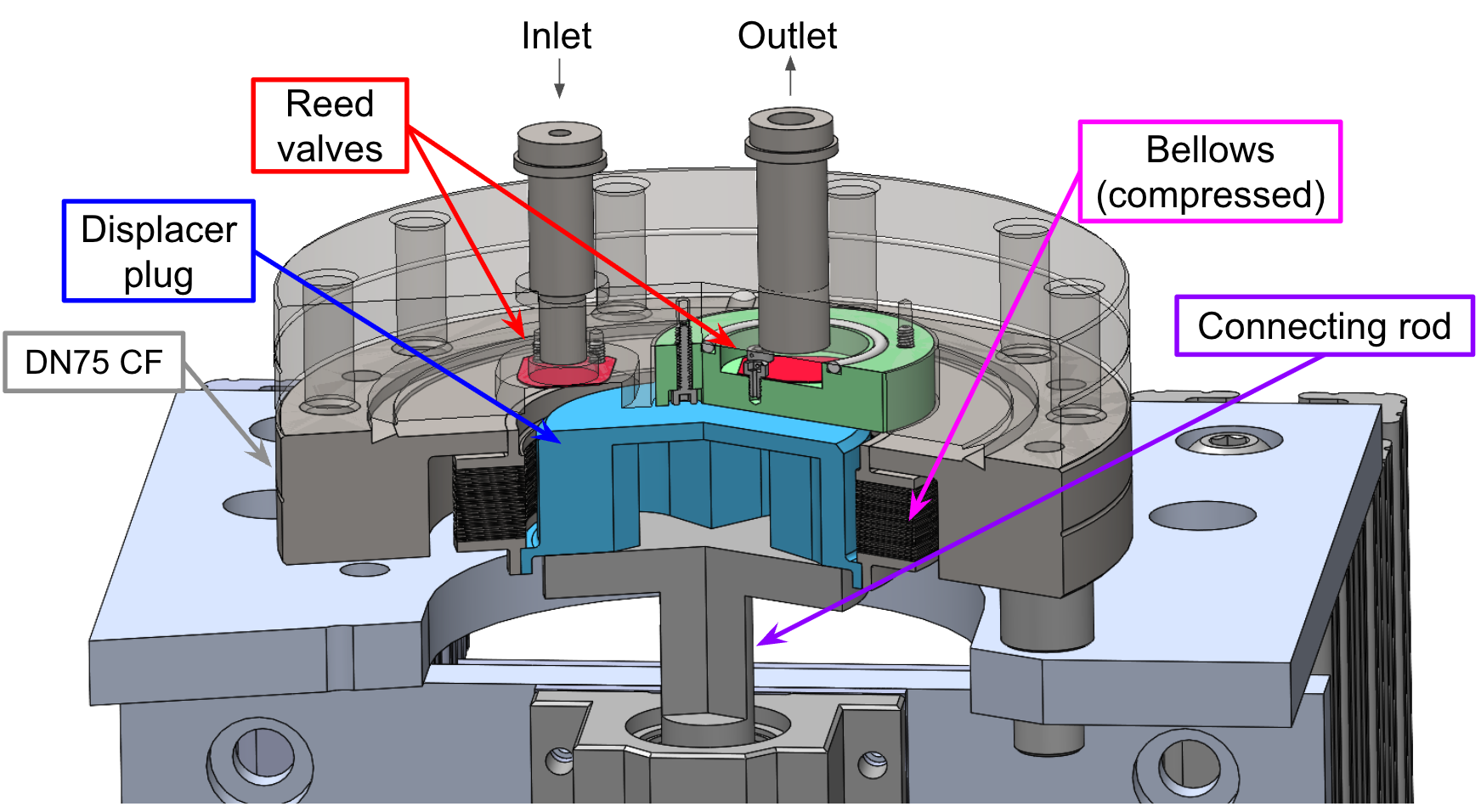}
  \caption{Cross section of the bellows pump CAD. Highlighted are the reed valves (red), displacer plug bottom (cyan), and inverting reed mount disc (green) sealed with a PTFE O-ring to the top flange for the output reed. The pump is cycled by a crankshaft (out of frame) attached via the connecting rod.}
  \label{fig:pump_crossection}
\end{figure*}

Recirculation pumps are essential components for these instruments, transporting radon from the emanation chamber into the ESC vessel with the carrier gas.
No commercial pump satisfies both low leak rates and low radon emanation, so a custom pump was built. 
The pump, shown in Fig.~\ref{fig:pump_crossection}, cycles a bellows with reed valves to regulate flow direction.
The top of the bellows is welded into a DN75 CF flange and a displacer plug is welded into the bottom to minimize volume at the top of the stroke.
The mating DN75 CF flange contains the reed-valves and the input and output ports.
The input reed is mounted in a shallow recess in the flange, allowing one way flow into the compression chamber.
The output reed is mounted on a stainless steel disc such that the reed flow direction is reversed. This disc is sealed (internally) to the top flange with a PTFE O-ring.
The bellows is actuated with a stepper motor turning a crankshaft to achieve the desired flow typically around $0.1-0.2~\text{SLM}$. The flow is determined with an inline mass flow meter (MKS G50A).
The first pump has been in continuous operation for over two years, however the bellows is susceptible to fatigue resulting in leaks.
The manufacturer (BellowsTech) only guarantees 3 million strokes ($\sim 2$ months at the current rate). 
Air getting into the system from a breach in the bellows could damage samples such as an expensive rare-gas purifier, so a magnetically coupled piston pump \cite{LePort2011,Brown2018} is in development for future systems.

\begin{figure*}[t]
\centering
  \includegraphics[width=0.9\textwidth]{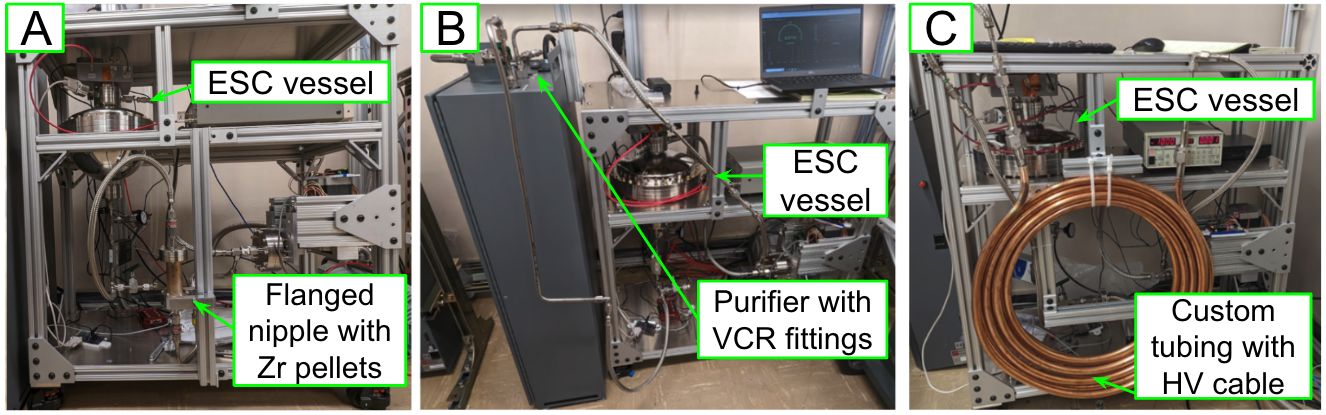}
  \caption{Examples of three different emanation vessels: A) DN63 CF nipple with Zr pellets, B) SAES rare gas purifier with VCR fittings, and C) 50 ft long copper tube with brazed VCR fittings for a HV cable.}
  \label{fig:sample_chambs}
\end{figure*}

The geometry of the emanation chamber depends on the sample being measured; three examples are shown in Fig.~\ref{fig:sample_chambs}. 
For small samples, an off-the-shelf CF nipple is used as the emanation chamber (Fig.~\ref{fig:sample_chambs}A). 
There are also parts of a xenon recirculation system that have ultra-high-vacuum compatible connectors which allow the sample to be directly connected in the recirculation loop (Fig.~\ref{fig:sample_chambs}B).
For large or odd-shaped samples, a custom chamber can be made (Fig.~\ref{fig:sample_chambs}C).
After closing the system, a leak test is performed with a Varian MD 30 Helium Leak Detector, which can detect leaks as low as $5 \times 10^{-10}$ Pa l/s. 
This minimum detection threshold corresponds to no more than $1$ radon atom ingress per year, assuming $15~\text{Bq/m}^3$ atmospheric concentration outside of the ESC recirculation volume.

\subsection{Radon Detection and Efficiencies}
\label{sec: rn_decay_eff}
\begin{figure*}
\centering
  \includegraphics[width=0.9\textwidth]{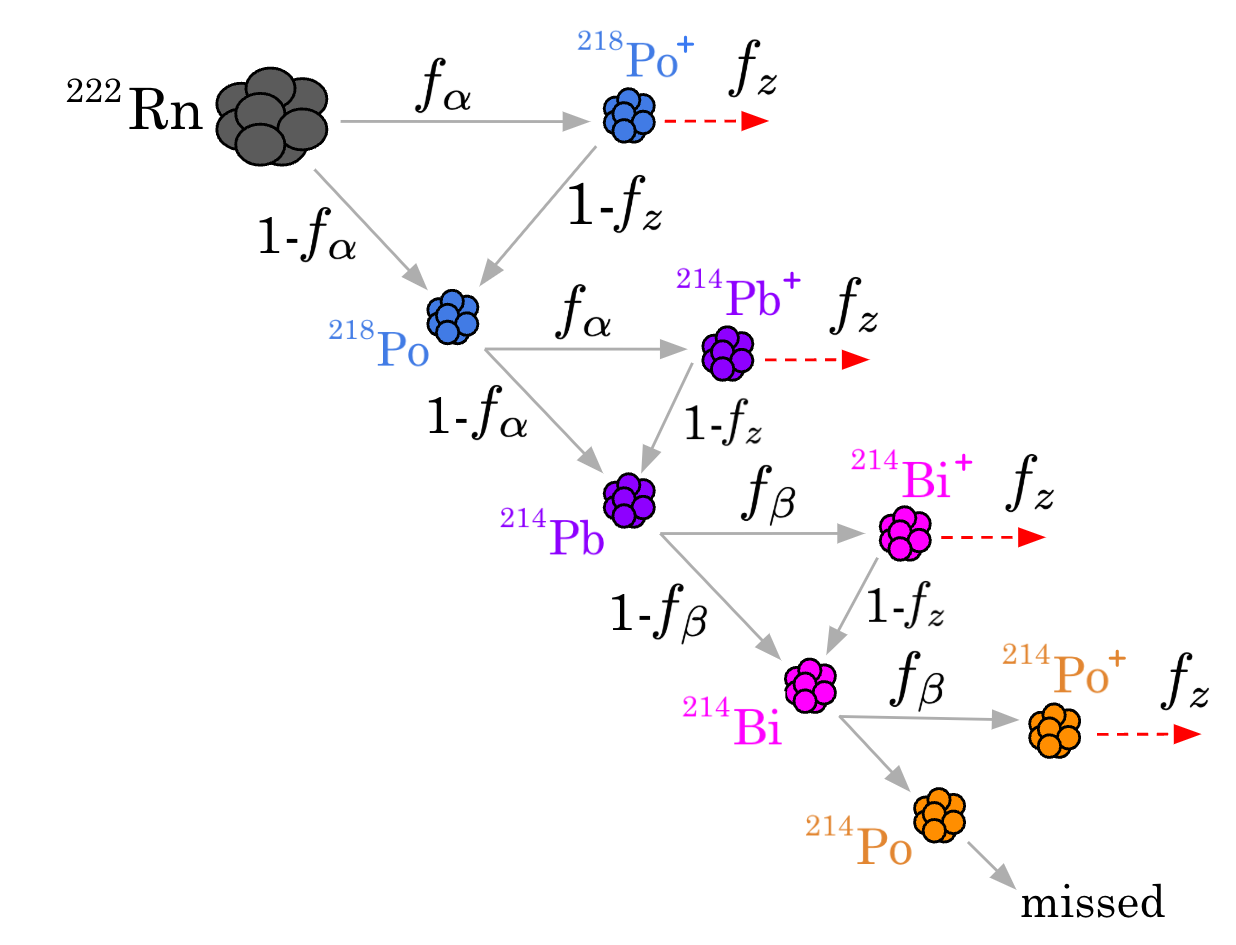}
  \caption{Diagram of $^{222}\text{Rn}$ collection probabilities inside the ESC. The value $f_\alpha$ ($f_\beta$) represents the positive ion fraction of an atom undergoing $\alpha$ ($\beta$) decay while $f_z$ represents the ion collection efficiency on the photodiode.}
  \label{fig:rn_decay_proc}
\end{figure*}

The overall detection efficiency of an ESC is the product of three quantities: the fraction of positive daughter ions, ion collection efficiency, and the photodiode's detection efficiency. 
The full chain of daughter ion collection probabilities is shown in Fig.~\ref{fig:rn_decay_proc}.
When $^{222}\text{Rn}$ $\alpha$-decays, the fraction $f_\alpha$ of ionized daughters, $^{218}\text{Po}^+$, is high, 88\% in air \cite{HOPKE1989299, PAGELKOPF20031057}, 50\% in LXe \cite{alphaion}.
All subsequent $\alpha$-decays are assumed to have the same $f_\alpha$.
Similarly for $\beta$-decays, the daughter ion fraction is denoted by $f_\beta$.
An ion in the ESC is then drifted to the Si photodiode with a probability $f_z$.
When progeny not attached to the diode undergo $\alpha$- or $\beta$-decay, they can also produce ions and be collected on the diode.
Once ions reach the Si photodiode, the detection efficiency of each subsequent $\alpha$ or $\beta$ decay particle entering the diode is 50\%. 
The detection of $^{214}\text{Po}$ is most critical for the $^{222}\text{Rn}$ analysis.

There is an expected difference between the count rates of $^{218}\text{Po}$ and $^{214}\text{Po}$ because of the less than unitary collection efficiency. 
If $^{222}\text{Rn}$ decays into neutral $^{218}\text{Po}$, it will not be collected on the Si photodiode, but there is a chance that its progeny will be ions and collected.
The ratio of $^{218}\text{Po}$ to $^{214}\text{Po}$ counts is called $\epsilon_{84}$ and is used as a diagnostic metric for our assays (see Sec. \ref{sec:measurements}).
Nevertheless, individual efficiency terms ($f_\alpha$, $f_\beta$, $f_z$) are not needed as the overall detection efficiency of $^{214}\text{Po}$ is measured with a calibration source, which is discussed in Sec.~\ref{sec:esc_calib}.

\subsection{Electronics and Data Acquisition}
The ESC electronics are custom, and the circuit design files are hosted on a public GitHub repository \cite{ESC_Electronics}.
To measure radon progeny, the Si photodiode is biased at -$1000~\text{Volts}$ to establish the drift field for ions in the ESC.
The photodiode is also operated with a $70~\text{V}$ reverse bias supplied via batteries floating at the HV-bias potential.
When radon daughters decay on the photodiode's surface, and if the $\alpha$ or $\beta$ particles enter the active silicon, they liberate charge proportional to the deposited energy. 
The freed charges are sensed on a blocking capacitor by a charge-sensitive preamplifier (Cremat CR110) that outputs a peak voltage $\mathcal{O}(100~\mathrm{mV}$) signal for a $6~\mathrm{MeV}$ $\alpha$, and return to baseline with a $140~\mathrm{\mu s}$ decay constant.
A second amplification stage is for voltage gain which increases the signal to $\sim 1~\mathrm{V}$ for the digitizer.

\begin{figure*}[t]
\centering
  \includegraphics[width=0.8\textwidth]{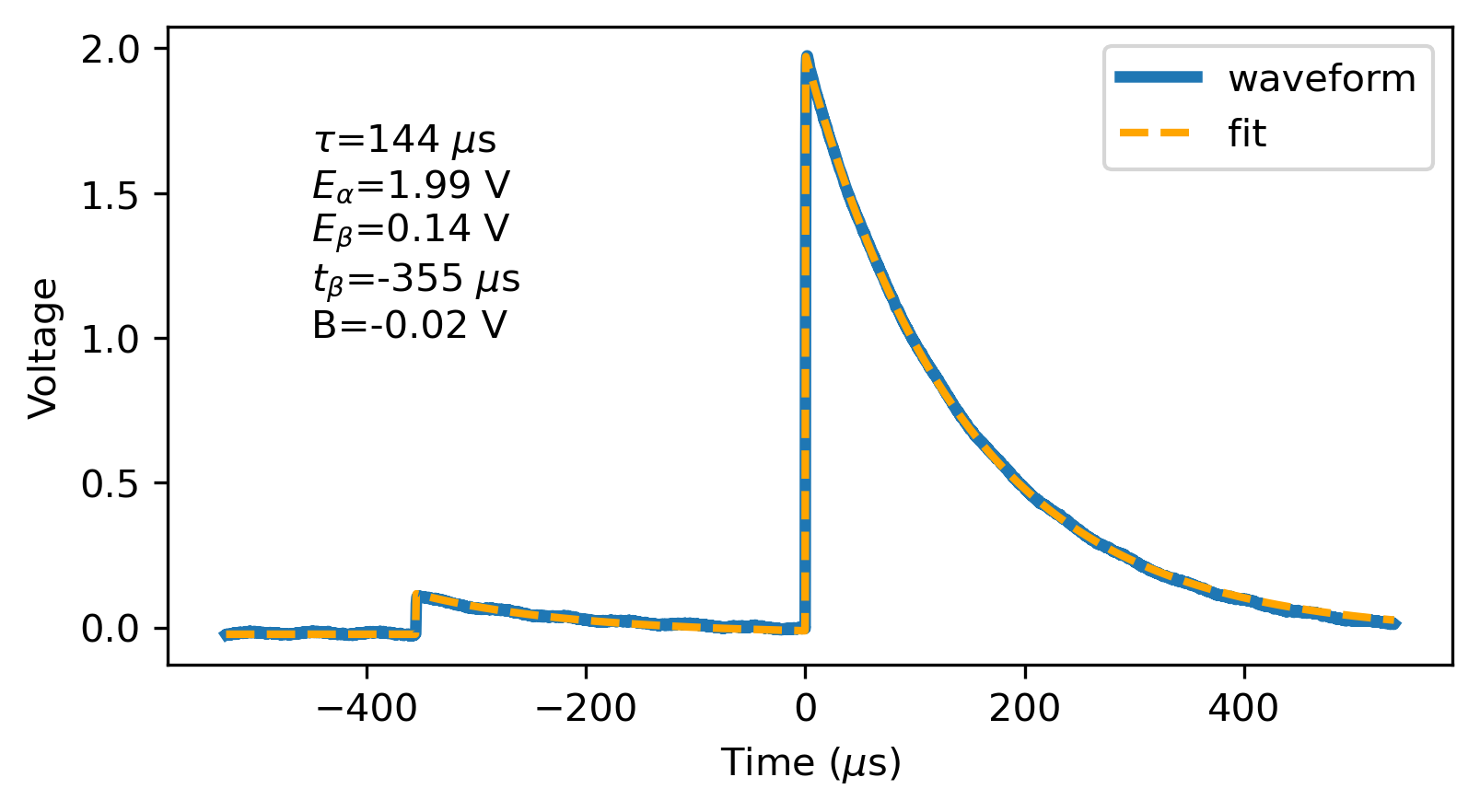}
  \caption{Waveform from a $^{214}$Bi - $^{214}$Po signal (blue) and the fit to the waveform (orange). Each signal peak is fit as a decaying exponential with decay constant $\tau$, baseline $B$, and amplitudes $E_\alpha$ and $E_\beta$. The second signal is assumed to be the $^{214}\mathrm{Po}$ $\alpha$ at $t=0$ (trigger), and the first is parameterized as $t_\beta$.}
  \label{fig:waveform_bi_po}
\end{figure*}

The amplified signal is digitized using a Digilent USB Oscilloscope (Analog Discovery Series 2 or 3) when voltages exceed a trigger threshold. The threshold is set below the $^{210}$Po $\alpha$-decay and above most $\beta$-decays ($\sim 0.2~\mathrm{V}$).
Waveforms are digitized at $4~\text{MHz}$ and saved $500~\mathrm{\mu s}$  pre- and post-trigger to allow adequate data to fit the baseline and decaying tail. 
This window has an efficiency for capturing both signals from the $^{214}$Bi-$^{214}$Po ($T_{1/2}=163~\mathrm{\mu s}$) of $\sim 88\%$.
The $^{214}$Bi $\beta$ is not used in the analysis, but is a nuisance for determining the baseline of the $^{214}\text{Po}$ $\alpha$.
An example waveform with a $^{214}\text{Bi}$ signal is shown in Fig~\ref{fig:waveform_bi_po}.
For typical assays, events are triggered at mHz rates. However, for calibration runs using a $62~\text{Bq}$ $^{226}\text{Ra}$ source, inline pile-up becomes an issue. 
The calibrations procedure (Sec.~\ref{sec:esc_calib}) has therefore been modified such that pileup events are negligible ($< 0.1\%$).

\section{PyDAn Analysis Software}
\label{sec:PyDAn}
A Python Data Analysis (PyDAn) framework was developed for this work.
PyDAn comprises three main classes: waveform input, energy calibration, and fitting of the Bateman equations \cite{bateman1910}. 
These classes are described in the following subsections.

\subsection{Data Processing}
The waveforms are fit offline assuming a $^{214}\text{Bi} - $$^{214}\mathrm{Po}$ like signal model (e.g. Fig.~\ref{fig:waveform_bi_po}), in which the primary $\alpha$ event is preceded by a smaller pulse from the $\beta$ event. 
This fitting strategy allows the baseline before the $\alpha$ event to be accurately fit without prior information on the type of event. 
The waveform model, which assumes the $\alpha$ peak is at the trigger location ($t=0$), has five free parameters: the voltage amplitude of each pulse ($E_\alpha$, $E_\beta$), the pre-trigger peak time ($t_\beta$), the decay constant of the charge-sensitive amplifier ($\tau$), and an overall baseline offset ($B$). 
Before the fit, the gradient of the waveform is calculated to find the number of peaks above a user-defined threshold. 
If a singular peak is found, $t_\beta$ is set to the first data point in the waveform (-$500~\mathrm{\mu s}$).
This allows for the possibility that the initial $\beta$-decay is prior to the capture window, and fits the baseline with the decaying exponential.
If there is no $\beta$-decay in the waveform the fit will find $E_\beta=0$, i.e. a horizontal line.
The saving and fitting of waveforms significantly improved the energy resolution at the cost of managing larger data-sets compared to the Multichannel Analyzers (MCA) used prior. 
Calibration runs produce $\sim 5~\text{GB}$ of waveform data during a typical two week data collection run.
After fitting, the $E_\alpha$ values are binned in energy and time to emulate the MCA behavior. 
The remaining analysis is then agnostic to the DAQ hardware.

\subsection{Energy Calibration and Detector Response}
\label{subsec:energy_calib}
\begin{figure*}
    \centering
    \includegraphics[width=0.85\linewidth]{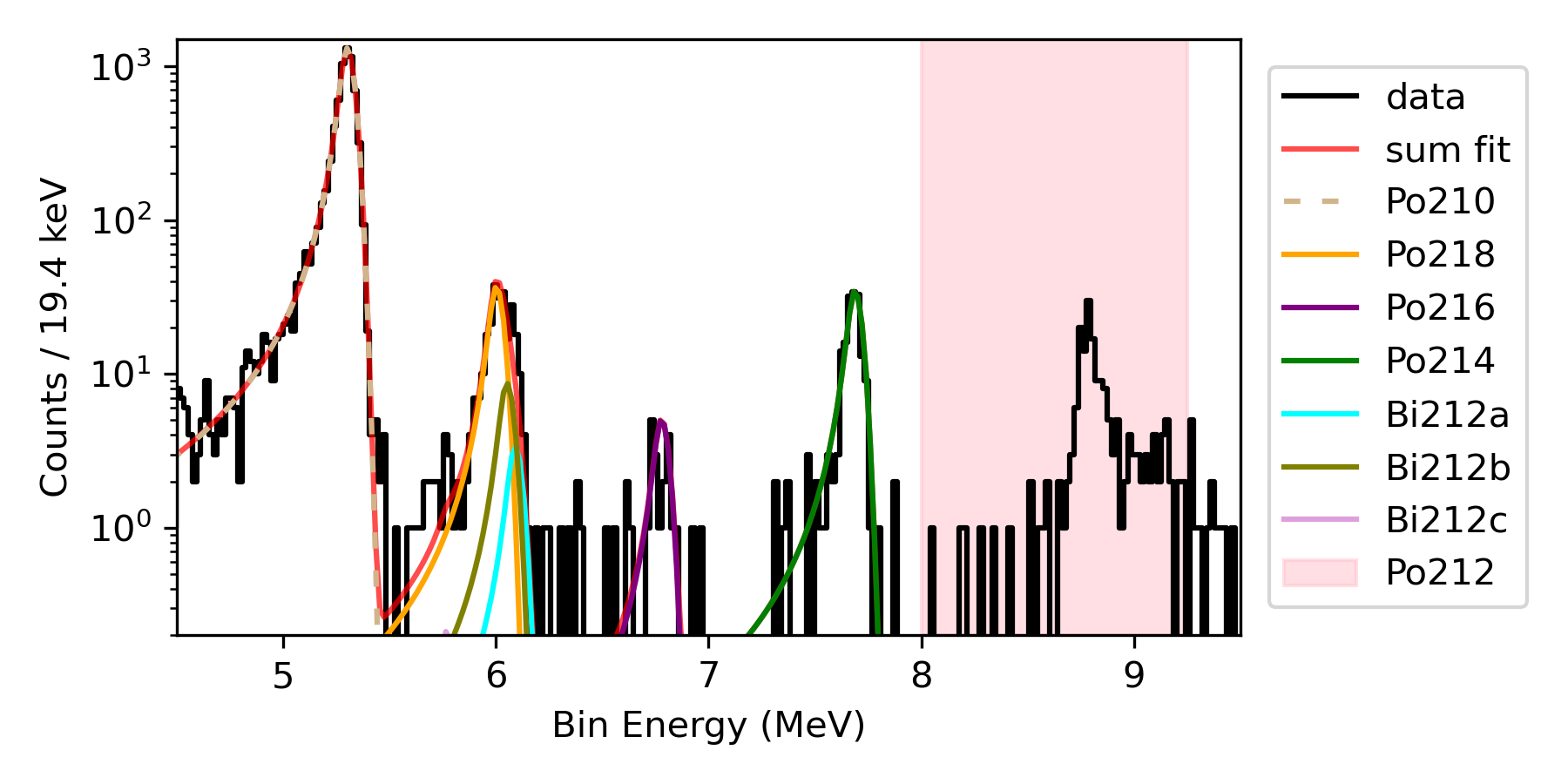}
    \caption{Histogram of a complex energy spectrum containing $^{222}\text{Rn}$ and $^{220}\text{Rn}$ progeny and PyDAn's fits of those $\alpha$ peaks. Spectrum is from the SAES PS4-MT50 purifier hot run. The red highlighted region (8-9.25 MeV) contains a $^{212}\text{Bi}+^{212}\text{Po}$ ($\beta+\alpha$) coincidence ($t_{1/2}=300~ns$) that is not resolved with our DAQ, creating an additional tail to the right of the peak. PyDAn associates all events in this region to $^{212}\text{Po}$ rather than fit the peak shape.}
    \label{fig:energy_fit}
\end{figure*}

A histogram of $E_\alpha$ is used to calibrate the charge electronics voltage to the deposited particle energy.
PyDAN automates the calibration process, allowing the user to assign two (or more) peaks in the spectrum to known $\alpha$-decays and performing a linear fit to convert $E_\alpha$ (voltage) to $\alpha$-energy. 
There are two parameters that describe the peak's
shape: a Gaussian width for the right half of the peak ($\sigma_g$), and the Lorentzian width on the left side of the peak ($\sigma_l$).
The Lorentzian shape provides a good approximation of the energy lost in the thin dead layer on the Si photodiode, as seen in Fig.~\ref{fig:energy_fit}. 
The output of the fit gives the energy calibration, the detector energy resolution, and the probability density functions (PDFs) for each $\alpha$-peak in the energy histogram. 
Figure~\ref{fig:energy_fit} shows the histogram of a sample containing peaks from both $^{220}\text{Rn}$ and $^{222}\text{Rn}$ decay series, and the fitted PDFs. 

\subsection{Fitting Initial Populations}
\label{sec:fit_populations}
\begin{figure*}
    \centering
    \includegraphics[width=0.8\linewidth]{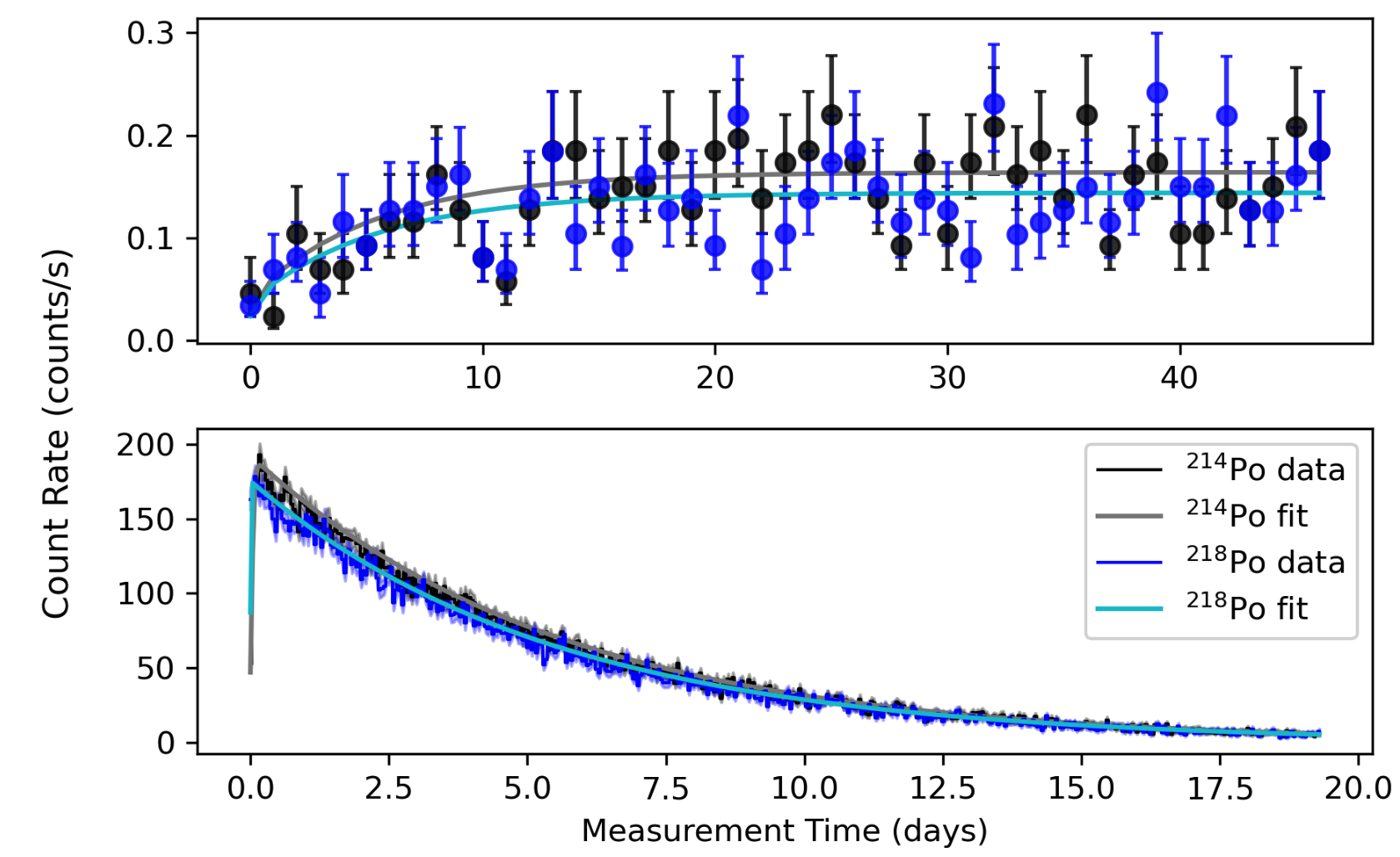}
    \caption{$^{214}\text{Po}$ and $^{218}\text{Po}$ count-rates versus time for an assay of ceramic beads (top, also see Table~\ref{tab:assays_list}) and a Pylon Rn-source injection (bottom, also see Section~\ref{sec:esc_calib}). The fits (cyan and gray curves) show the count rates the model predicts in each time bin. For the ceramic beads, the count rates build up to steady state as the sample emanates $^{222}\text{Rn}$, and then the count rate is scaled by the detection efficiency and background subtracted to get the final rate of $290~\mathrm{\mu Bq}$. For the calibration run, the system is injected with $377~\mathrm{mBq}$ $^{222}\text{Rn}$ which decays during the measurement, and the count rates decrease over time.}
    \label{fig:pyl_ceram}
\end{figure*}

After calibration, the time-binned energy histograms are fit to the solutions of the Bateman equation.
PyDAn uses the NumPy library \cite{numpy} to solve the system of differential equations with the formalism described in \cite{Levy2018}.
Briefly, the decay series is described by a set of differential equations 
\begin{equation}
\begin{split}
N_1'(t) = -\lambda_1 N_1(t) \\
N_2'(t) = \lambda_1 N_1(t) -\lambda_2 N_2(t) \\
\cdots \\
N_n'(t) = \lambda_{n-1} N_{n-1}(t) -\lambda_n N_n(t), \\
\end{split}
\end{equation}
where $\lambda_{n}$ are the decay time constants, and $N_n$ the populations.
This series can be expressed as $\matr{N}'(t) = \matr{A} \matr{N}(t)$, where
\[
\matr{A} = 
\begin{bmatrix}
-\lambda_1 & 0 & \cdots & 0 \\
\lambda_1 & -\lambda_2 & \cdots & 0 \\
\cdots & \cdots & \cdots & \cdots \\
0 & \cdots & \lambda_{n-1} & -\lambda_{n} \\
\end{bmatrix} .
\]
PyDAn populates the decay series data into $\matr{A}$ for each series with hard-coded values from the ENSDF database \cite{ensdf}.
Using NumPy, the eigenvector and eigenvalues are found such that $\matr{A} = \matr{V} \matr{\Lambda} \matr{V}^{-1}$.
The general solution for the populations $\matr{N}(t)$ at a later time $t$ is then given by:
\begin{equation}
\label{eq:N}
\matr{N}(t) = \matr{V} e^{\matr{\Lambda}t} \matr{V^{-1}} \matr{N}_0 .
\end{equation}
There is a transient period of a few hours at the beginning of each run where the populations of the progeny following radon will stabilize. 
These initial populations are not of interest for assay measurements, and PyDAn sets them to zero. This is typically true at the start of a run; otherwise the first few hours of the data can be masked out. 
Either way, this short stabilization time does not affect the fit at the long time scales of our assays. 
From Eq. \ref{eq:N}, the expected number of $\alpha$ decay counts per time bin ($\Delta \matr{N}(t_i)$) can be obtained from the remaining free parameters: the initial populations of $^{226}\text{Ra}$, $^{222}\text{Rn}$, and $^{232}\text{Th}$ (the $^{232}\text{Th}$ chain is assumed to be in equilibrium up to $^{224}\text{Ra})$.
The $\alpha$-decay counts are multiplied by their unique PDFs ($PDF_j$) found in Sec. \ref{subsec:energy_calib} to get an energy and time binned model ($M_{i,j}$), a 2-Dimensional matrix that is fit to the data.
The fit is performed by minimizing the Negative Log Likelihood (NLL) cost function:
\begin{equation}
\label{eq:NLL}
\text{NLL} = \sum_{i,j} M_{i,j} - D_{i,j} \cdot \log (M_{i.j}) ,
\end{equation}
where $D$ is the observed number of counts in each energy-time bin.
The model ($M_{i,j}$) is defined as:
\begin{equation}
\label{eq:Mij}
M_{i,j} = PDF_j \cdot \Delta N_j(t_i) .
\end{equation}
Two additional free parameters are created to allow the ratios of $^{218}\text{Po}$ to $^{214}\text{Po}$ counts and $^{216}\text{Po}$ to $^{212}\text{Po}$ counts to be less than one due to their differing collection efficiencies, $\epsilon_{84}$ and $\epsilon_{62}$ respectively (see Sec. \ref{sec: rn_decay_eff}). 
This means that $^{214}\text{Po}$ and $^{212}\text{Po}$ counts constrain $\matr{N}_0$ whereas $\epsilon_{84}$ and $\epsilon_{62}$ provide mostly diagnostic information about the ion collection efficiency.
When $\epsilon_{84}$ is less than $\sim 0.7$, the assay is suspect and the sample is conditioned and re-measured; the concern being the ESC detection efficiency is not known as the ions are not being collected at the nominal rates. 
These assays have only occurred with plastic and rubber samples where contaminants can neutralize ions.
The initial population found for $^{226}\text{Ra}$ is the population that supports the $^{222}\text{Rn}$ emanation, the quantity of interest for an assay. 
The initial population found for $^{222}\text{Rn}$ is the number of atoms initially trapped in the system, which is used for the calibration process.
The number of $^{214}\text{Po}$ and $^{218}\text{Po}$ counts and the resulting fits for an active sample  are shown in Fig. \ref{fig:pyl_ceram} (top).

\section{Measurement Procedure and Results} 
\label{sec:measurements}
There are three measurements required to interpret the fit results as an emanation measurement: detection efficiency calibration, measurement of the background, and measurement of the sample. 
This section first describes the method for performing a calibration run, then the instrument background, and concludes with a table of emanation rates for a number of assayed samples.

\subsection{ESC Calibration}
\label{sec:esc_calib}
Detection efficiency of the ESCs is performed with a $62 \pm 2.2~\text{Bq}$ $^{226}\text{Ra}$ RNC flow-through source manufactured by Pylon Electronics Inc~\cite{pylon}. 
The source is evacuated of all $^{222}\text{Rn}$ and flushed with the carrier gas.
It is then filled with carrier gas to 1 atm and isolated. 
The source is allowed to emanate for a fixed duration, typically 1 hour, to build up a known $^{222}\text{Rn}$ population. 
The isolation valves are then opened to the ESC vessel and a carrier gas is pushed through the source to the evacuated ESC at $\sim 4~\text{SLM}$.
This transports the $^{222}\text{Rn}$ while filling the system to 1~atm.
The amount of injected $^{222}\text{Rn}$ is given by
\begin{equation}
\label{eq:Arn}
A_{\mathrm{Rn}} = \frac{\lambda_{\mathrm{Rn}}}{\lambda_{\mathrm{Rn}}-\lambda_{\mathrm{Ra}}} A_{\mathrm{Ra}} (e^{-\lambda_{\mathrm{Ra}}t}-e^{-\lambda_{\mathrm{Rn}}t}) ,
\end{equation}
where $\lambda_{\mathrm{Rn}}$ and $\lambda_{\mathrm{Ra}}$ are the decay constants of $^{222}\mathrm{Rn}$ and $^{226}\mathrm{Ra}$,  $A_\text{Rn}$ and $A_\text{Ra}$ the activity rates, and $t$ is the time from the start of the source emanation to the end of the ESC fill.
The calibration is performed without the recirculation loop so that all $^{222}\mathrm{Rn}$ is inside the ESC vessel.

The detection efficiency of the ESC is determined by measuring $N_0$ for $^{222}\text{Rn}$ found by PyDAn, then dividing by the number of atoms injected ($N = A_{\mathrm{Rn}} / \lambda_{\mathrm{Rn}}$).
A typical Pylon-injection dataset with fits is shown in Fig.~\ref{fig:pyl_ceram} (bottom).
Each instrument is calibrated with each carrier gas, and the overall detection efficiencies have ranged from $0.35$ to $0.45$, depending on the combination of carrier gas and ESC used.
When keeping these parameters constant, the variation between measurements was less than 2\% for three distinct runs.
The uncertainty of the detection efficiency is dominated by the Pylon source activity uncertainty, at 3.2\%. 
The efficiency for $^{220}\text{Rn}$ progeny has not yet been determined and is a more difficult problem left for future studies. 
In this work, the $^{220}\text{Rn}$ chain is not considered further. 

The last efficiency term to account for is the volume-sharing factor, which is the ratio of the ESC vessel volume to total system volume. 
For example, the empty CF nipple emanation chamber used for small sample assays has a ratio of 0.92, however this factor is estimated for each measurement.
This accounts for the inability of the ESC to collect  progeny that decay in the volume outside of the ESC vessel. 
The fit value of the initial population of $^{226}\text{Ra}$ multiplied by the corresponding decay constant is the detected emanation rate.
This quantity is then scaled by the detection efficiency and volume fraction to determine the actual emanation rate.
This scaling process is used for all emanation measurements, background and sample runs. 

\subsection{Background Measurements}
Instrument backgrounds need to be subtracted from sample measurements to determine the emanation rate of the sample. 
This measurement is performed with the entire recirculation system in a configuration as close to the sample measurement state as possible. 
For samples that fit into emanation chambers (see Fig.~\ref{fig:sample_chambs}A/C), the background is measured with the empty chamber. 
For inline objects like the xenon purifier (see Fig.~\ref{fig:sample_chambs}B), the background measurement is performed using a bypass loop with a valve that can be opened while the purifier inlet and outlet valves are closed. 
Many assays require reconfiguring the recirculation loop plumbing to accommodate the sample, so no two background runs are equivalent. Each sample gets a dedicated background measurement. 
However, the instrument backgrounds have been measured in nine configurations, with an average rate of $197~\mathrm{\mu Bq}$ and an average statistical error of $27~\mathrm{\mu Bq}$.
Not every measurement was performed to the same level of sensitivity due to needs of the particular sample, yet the standard deviation of those nine unrelated background measurements is $23~\mathrm{\mu Bq}$.
Our conclusion was that the major source(s) of backgrounds were from the common components of the instrument like the ESC vessel or pump, and that these measurements are repeatable.

\subsection{Assay Results}
Table~\ref{tab:assays_list} gives assay results of several samples, some potentially useful for future low-background experiments.
For each, a background measurement was performed and subtracted in the final result while the uncertainties for sample and background were added in quadrature. 
With long measurement times, up to 4 weeks, measurement uncertainties of $\sim 20~\mathrm{\mu Bq}$ at one standard deviation have been achieved (see Beryllium Copper springs in Table~\ref{tab:assays_list}).
\begin{table}[ht]
    \centering
    \caption{Background subtracted radon emanation measurements with either a one-standard deviation error or a 90\% CL (confidence level) as described in Ref. \cite{Tsang2023}.The Sample and Background columns describe the measured rates with and without the sample present. Counting time for each is given in parentheses below the measured value. $^{*}$The heaters were modified to run at 550 \textdegree{C}.}
    \small
    \begin{threeparttable}[t]
        \begin{tabular}{C{0.24\linewidth}|C{0.2\linewidth}|C{0.1\linewidth}|c|C{0.12\linewidth}}
        
Sample & Description & Background [$\mu$Bq] & Sample [$\mu$Bq] & Emanation [$\mu$Bq] \\
\hline\hline
Beryllium Copper springs & 10 springs O.D 0.5~in and 4~in length. MFG Century Spring, PN 10693CS & \makecell{$153 \pm 20$ \\ (21 days)} & \makecell{$195 \pm 20$ \\ (29 days)} & $42 \pm 29$ \\ 
\hline
GetterMax 133 & 357~g copper coated beads. MFG Research Catalysts & \makecell{$210 \pm 25$ \\ (21 days)} & \makecell{$2050 \pm 145$ \\ (12 days)} & $1840 \pm 150$ \\
\hline
\multirow{2}{*}{SAES PS4-MT3} & 
Purifier assembly containing 500~g of ST707 & 
\makecell{$211 \pm 28$ \\ (19 days)} & 
\makecell{heaters ON$^{*}$: $221 \pm 21$ \\ (29 days)} &  $< 70$ \\
\cline{4-5}
& & & \makecell{heaters OFF: $220 \pm 26$ \\ (20 days)} & $< 73$ \\
\hline
\multirow{2}{*}{\makecell{\\SAES PS4-MT50-R-535}} & 
Purifier assembly containing \mbox{$\sim 4.4~\mathrm{kg}$} ST707 & 
\makecell{$193 \pm 23$ \\ (22 days)} & 
\makecell{heaters ON: $622 \pm 56$ \\ (19 days)} & $428 \pm 61$ \\
\cline{4-5}
& & & \makecell{heaters OFF: $369 \pm 42$ \\ (18 days)} & $176 \pm 48$ \\
\hline
Dielectric Science HV cable & Two 5~m pieces Polyethylene HV cable stripped of braid and jacket procured by PNNL, PN 2353 & \makecell{$175 \pm 22$ \\ (31 days)} & \makecell{$206 \pm 25$ \\ (27 days)} & $< 81$ \\
\hline
Ceramic beads & Alumina vacuum insulator beads totaling 200~g. MDC CB-1 680600 & \makecell{$130 \pm 17$ \\ (21 days)} & \makecell{$420 \pm 24$ \\ (47 days)} & $290 \pm 29$ \\
\hline
Zirconium pellets & ALB Materials (99.95\% 2x2~mm cylinder) totaling 438~g & \makecell{$214 \pm 27$ \\ (20 days)} & \makecell{$381 \pm 36$ \\ (20 days)} & $168 \pm 45$ \\
\hline
        \end{tabular}
    \end{threeparttable}
    \vspace{-10pt}
    \label{tab:assays_list}
\end{table}

The quoted uncertainties in the emanation results are the combination of three sources added in quadrature. 
First is the uncertainty from the calibration measurement, which includes procedural uncertainty (associated with timing the injection of radon), statistical uncertainty (one-standard deviation as reported by \textit{iminuit}), and the Pylon source activity uncertainty (3.2\%). The combination of these factors is 3.7\%.
Second is the uncertainty in the volume sharing between the ESC vessel and the recirculation plumbing; this value varies depending on the sample, but is typically less than 1\%. 
Third is the statistical uncertainty of the assay measurement, which is obtained in the analysis from PyDAn's fit using \textit{iminuit} with one-standard deviation error.
For measurements with a low count rate, the overall uncertainty is dominated by this third factor.

\section{Conclusion}
In many low-background, low rate-experiments, $^{222}\text{Rn}$ and its daughters are a significant source of background that can limit sensitivity to detecting new physics like $0\nu\beta\beta$ and WIMP particle interactions.
The stringent radon requirements for rare-event searches demand material screening with ultra-sensitive radon detectors. 
This work describes a highly sensitive technique for measuring $^{222}\text{Rn}$ emanation using a recirculating system with an ESC detector. 
To perform the data analysis and estimate the population of the radon emanation supporting species, a data analysis framework, called PyDAn, was developed.
The instruments' detection efficiencies have been determined with a commercial $^{222}\text{Rn}$ source, the procedure of which is discussed. 
Finally, the $^{222}\text{Rn}$ emanation results are presented for a number of samples. 
These instruments are capable of measuring $^{222}\text{Rn}$ emanation rates as low as $20~\text{$\mu$Bq}$ with runs times $\sim 4~\mathrm{weeks}$ long.
This sensitivity is, to our knowledge, world leading for material assay.
Future efforts will aim to lower the backgrounds of the instruments to improve sensitivity further.
Operations are being moved to an ISO5 clean-room where particulate levels are lower.
Additionally, studies of citric-acid passivation, and nitric-acid etching will be performed on the internal surfaces in an attempt to further reduce instrument backgrounds.

\section{Acknowledgments}
This work was supported in part by Laboratory Directed Research and Development (LDRD) programs at Brookhaven National Laboratory (BNL), Lawrence Livermore National Laboratory (LLNL), Pacific Northwest National Laboratory (PNNL), and SLAC National Accelerator Laboratory. The authors gratefully acknowledge support for nEXO from the Office of Nuclear Physics within DOE’s Office of Science under grants/contracts DE-AC02-76SF00515, DE-FG02-01ER41166, DE-SC002305, DE-FG02-93ER40789, DE-SC0021388, DE-SC0012704, DE-AC52-07NA27344, DE‐SC0017970, DE-AC05-76RL01830, DE-SC0012654, DE-SC0021383, DE-SC0014517, DE-SC0024666, DE-SC0020509, DE-SC0024677 and support by the US National Science Foundation grants NSF PHY-2111213 and NSF PHY-2011948; from NSERC SAPPJ-2022-00021, CFI 39881, FRQNT 2019-NC-255821, and the CFREF Arthur B. McDonald Canadian Astroparticle Physics Research Institute in Canada; from IBS-R016-D1 in South Korea; and from CAS in China.

\bibliography{bib}

\end{document}